\advance\voffset by 1mm
\advance\hoffset by -7mm
\makeatletter
\@addtoreset{equation}{section}
\makeatother

\renewcommand{\title}[1]{\null\vspace{25mm}

\noindent{\Large{\bf #1}}\vspace{10mm}  

\noindent {\large By }}
\newcommand{\authors}[1]{\noindent{\large #1}\vspace{3mm}

}
\newcommand{\address}[1]{\noindent #1\vspace{5mm}

}

\documentstyle[11pt]{article}

\newtheorem{th}{Theorem} 
\newtheorem{lm}{Lemma}
\newtheorem{df}{Definition}  
\newtheorem{pr}{Proposition}
\newtheorem{as}{Assumption}
\newcommand{\bth}{\begin{th}}
\newcommand{\eth}{\end{th}}
\newcommand{\blm}{\begin{lm}}  
\newcommand{\elm}{\end{lm}}
\newcommand{\bdf}{\begin{df}} 
\newcommand{\edf}{\end{df}} 
\newcommand{\bpr}{\begin{pr}}
\newcommand{\epr}{\end{pr}}
\newcommand{\bas}{\begin{as}}
\newcommand{\eas}{\end{as}}
\newcommand{\bpf}{\noindent {\bf Proof:} }
\newcommand{\epf}{$\bullet$\par\vspace{1.8mm}\noindent}
\newcommand{\bit}{\begin{itemize}}
\newcommand{\eit}{\end{itemize}\par\noindent}
\newcommand{\beq}{\begin{equation}} 
\newcommand{\eeq}{\end{equation}\par\noindent}
\newcommand{\beqa}{\begin{eqnarray*}}
\newcommand{\eeqa}{\end{eqnarray*}\par\noindent}
\newcommand{\beqn}{\begin{eqnarray}}
\newcommand{\eeqn}{\end{eqnarray}\par\noindent}
\newcommand{\la}{\lambda} 
\newcommand{\La}{\Lambda}
\newcommand{\si}{\sigma} 
\newcommand{\Si}{\Sigma}
\newcommand{\DL}{\Delta\Lambda}
\newcommand{\vlp}{\varphi_\lambda(p)}

\begin{document}

\newfont{\srm}{cmr8}
\newfont{\sit}{cmti8}
\newfont{\sdub}{msym7} 
\newfont{\dub}{msym7 scaled\magstep1}
\newfont{\ldub}{msym9 scaled\magstep1}

\newfam\msbfam
\font\tenmsb=msbm10			\textfont\msbfam=\tenmsb
\font\sevenmsb=msbm7		\scriptfont\msbfam=\sevenmsb
\font\fivemsb=msbm5			\scriptscriptfont\msbfam=\fivemsb
\def\Bbb{\fam\msbfam \tenmsb}

\def\insim{\, \raise-5 pt\hbox{$\buildrel\hbox{\hbox{$\sim$ \hskip - 12 truept \raise
5pt\hbox{$\scriptstyle \in$}}}\over{\scriptstyle\,\,}\ $}}
\def\duni{\raise-3 pt\hbox{$\buildrel\hbox{\hbox{$\bigcirc$ \hskip - 13 truept
\raise -0.1pt\hbox{$\bf \cup$}}}\over{\scriptstyle\,\,} $}\, }
\def\dunia{\raise-7 pt\hbox{$\buildrel\hbox{\hbox{$\bigcirc$ \hskip - 13.0 truept 
\raise -0.1pt\hbox{$\bf \cup$}\hskip - 6.0 truept\raise-6.5pt\hbox{$\scriptstyle
a$}}}\over{\scriptstyle\,\,} $}\,  \hskip 1.5pt}
\def\dsuma{\raise-7 pt\hbox{$\buildrel\hbox{\hbox{$\bigcirc$ \hskip - 13.7 truept
\raise -0pt\hbox{$\bf +$}\hskip - 6.5 truept\raise-6.5pt\hbox{$\scriptstyle
a$}}}\over{\scriptstyle\,\,} $}\, \hskip 1pt}
\def\sdsuma{\raise-5.1 pt\hbox{$\buildrel\hbox{\hbox{$\scriptstyle\bigcirc$ \hskip -
11.2 truept \raise -0pt\hbox{$\scriptstyle\bf +$}\hskip - 5.3
truept\raise-4.8pt\hbox{$\scriptscriptstyle a$}}}\over{\scriptscriptstyle\,} $}\,
\hskip 0.7pt}
\def\nonexist{\raise-8 pt\hbox{$\buildrel\hbox{\hbox{$\exists$ \hskip - 9.0 truept 
\raise -0.1pt\hbox{$|$}\hskip - 6.0 truept\raise-6.5pt\hbox}}\over{\scriptstyle\,\,} $}\hskip 8pt}
 
\title{Classical Representations for Quantum-like Systems
through an Axiomatics for Context Dependence.}  
\authors{Bob Coecke} 
\address{FUND-DWIS, Free University of Brussels,\\ Pleinlaan 2,
B-1050 Brussels, Belgium;\\ bocoecke@vub.ac.be}
\par
\vspace{-7.8cm}
\par
\noindent
This paper has been published in {\it Helv. Phys. Acta} {\bf 70}, 442 (1997).
\par
\vspace{8.8cm}
\par
\noindent
We introduce a definition for a 'hidden measurement system', i.e., a
physical entity for which there exist:
(i) 'a set of non-contextual states of the entity
under study' and 
(ii) 'a set of states of the measurement context',
and which are such that all uncertainties are due to a lack of knowledge on the
actual state of the measurement context.  First we
identify an explicit criterion that enables us to verify whether a given
hidden measurement system is a representation of a given couple
$\Si,{\cal E}$ consisting of a set of states $\Si$ and a set of measurements 
${\cal E}$ ($=$ measurement system). Then we prove for every
measurement system that there exists at least one representation
as a hidden measurement system with $[0,1]$ as set of states of the measurement context. Thus, we
can apply this definition of a hidden measurement system to impose an axiomatics for context
dependence.  We show that in this way we always find classical representations (hidden
measurement representations) for general non-classical entities (e.g. quantum entities).
 
\section{Introduction.}    
    
In \cite{aer86}, Aerts 
introduced the 'hidden measurement approach' to quantum mechanics. He considered the quantum
state as a complete representation of the entity under study, but he allowed a lack of
knowledge on the interaction of the entity with its measurement
context during the measurement. This idea can also be put forward as follows: {\it with every
quantum measurement corresponds a collection of classical measurements (called hidden
measurements), and there exists a lack of knowledge concerning which measurement is actually
performed}\footnote{For
a general physical and philosophical background of the idea of hidden measurements we refer to
\cite{aer86,aer94,aer95c,thesisbob}.}. Explicit 'hidden measurement
models' have been introduced for some 'typical' quantum systems (see 
\cite{aer86}, \cite{aer91}, \cite{aer95c},
\cite{coe95c},
\cite{thesisbob} and \cite{coehpa}).  

In this paper, we apply these idea's within a much more general framework. In stead of only
supposing the existence of a set of states for the physical entity (denoted by
$\Sigma$), we also suppose the existence of a set of states of the measurement context
(denoted by $\La$) which corresponds with the collection of hidden measurements. For an
as general as possible class of systems defined by a set
$\Si$ of states and a set
${\cal E}$ of measurements (called 'measurement systems' and abbreviated as m.s$.$) we will
prove that there exists an equivalent representation as a 'hidden measurement system' (abbreviated as
h.m.s$.$) such that the
probabilities that occur are due to a lack of knowledge on the actual state of the
measurement context.   
In this way we find for every m.s$.$, and thus also for quantum
mechanics, a classical representation as a h.m.s.

In section
\ref{subsec:comp} we illustrate how an additional structure on the m.s$.$
(for example, the geometric structure of quantum mechanics) can be induced on the h.m.s$.$ in
a natural way.  Thus, the classical representations that we consider respect the
symmetries of the given entity. We also identify  the criterion that enables us to verify whether a given
h.m.s$.$ is a representation of a given m.s$.$ (see section 4.2).  Such a criterion is
an essential tool for any further study that uses this 'hidden measurement axiomatics' for
context dependence.  
In \cite{thesisbob} 
and \cite{zoology} we have build a complete classification of all possible
h.m.s.-representations for a given quantum m.s., starting from this criterion.  
 
For a general definition of the basic mathematical objects that are
used in this paper ($\si$-fields,
$\si$-morphisms, probability measures, measurable functions etc...) we refer to
\cite{bir40} and \cite{sik69}.  We mention that from a mathematical point of view, the
representation that we introduce in this paper coincides sometimes with Gudder's proof on
the existence for contextual hidden variable representations\footnote{For the debate on this
kind of representations we refer to \cite{gud68},
\cite{jau63} and \cite{neu32}.} of systems described by
orthomodular lattices (see \cite{gud70}).  
A first  
theorem on the existence of a hidden measurement
representation for finite dimensional quantum mechanics was contained in
\cite{aer86}. A generalization of this theorem to more general
finite dimensional entities can be found in \cite{aer94}.  The specific case of mixed states
was considered in \cite{coe95a}, and the general proof for the existence of a hidden
measurement representation for infinite dimensional entities can be found in
\cite{coe95d}. Finally, we remark that the
results presented in this paper (except for section \ref{subsec:comp})
where made known in \cite{thesisbob}.  
 
\section{Assumptions of the approach.}

In this section we consider a situation when there is a lack 
of knowledge concerning the interaction of the entity under
study with its measurement context, i.e$.$, when the state\footnote{We
exclude the situation of a lack of knowledge concerning the state, i.e$.$, if we write
'state', we mean 'pure state'. For a well-founded definition of state
we refer to \cite{pir76}.} of the entity does not determine the outcome anymore. In such a
case,  when we perform a measurement $e$ on an entity in a state $p$, we might even be lucky
if we manage to find a formalizable statistical regime in the occurring outcomes. As a
consequence, a general theoretical treatment of these measurements is a priori not possible. 
Nevertheless, after stating a few reasonable assumptions, it is possible to construct a
framework to study these situations:    
\bas 
There exists\footnote{We
remark that 'existence' is not equivalent with 'knowledge'. Thus, we don't have to know
the set of possible descriptions of the measurement context.} a set of possible descriptions of 
the
measurement context on the precise time that we decide to perform the measurement, i.e., there
exists a set of 'relevant' parameters for the measurement context.  We call this set of relevant
parameters the 'states of the measurement context'.   
\eas 
\bas
The result of a measurement, which is the result of the interaction
between the entity and the measurement context, is completely determined by the state
of the entity and the state of the measurement context, i.e$.$, there is a 'deterministic
dependence' on the initial conditions.
\eas
\bas
There exists a statistical description for the
relative frequency of occurrence of the states of the measurement context during the
measurement.     
\eas
We suppose that all these assumptions are fulfilled. In the next sections, we will
denote the set of states of the measurement context as $\La$.
For a fixed state of the measurement context $\lambda\in\La$, the measurement process is
strictly classical\footnote{We use 'strictly classical' in stead of 'classical' since we
exclude the situations of unstable equilibrium that occur in most classical theories.} (because of
the deterministic dependence), and thus, for every such strictly classical {\it hidden
measurement} there exists a strictly classical observable:  
\beq 
\varphi_\la :\Si\rightarrow O_e 
\eeq
Where $\Sigma$ is the set of states
of the physical entity and $O_e$ is the set of possible outcomes of measurement $e$. Thus, we
have the following set of strictly classical observables that correspond with the different
possible states of the measurement context: 
\beq
\Phi_\La=\{\varphi_\la | \la\in\La\}
\eeq
Since there exists a relative frequency of occurrence for states of the measurement context, 
there exists
a probability measure: 
\beq
\mu_\La : {\cal B}_\La\rightarrow [0,1]
\eeq  
Where ${\cal B}_\La$ is a $\si$-field of
subsets of $\La$. Thus, we are able to compute a probability defined on subsets of the set of
outcomes, for every given initial state, i.e$.$, we obtain an 'outcome probability' for every
measurement
$e$ on the entity in a state $p$: 
\beq
P_{p,e}:{\cal B}_e\rightarrow [0,1]
\eeq
Where ${\cal B}_e$ is a $\si$-field of subsets of $O_e$. In fact, we have summarized, and
represented, the 'unknown but relevant information' of the measurement process (i.e$.$, all
possible interactions during the measurement, for all possible initial states), in a couple
consisting in: a set of strictly classical observables $\Phi_\La$
and, a probability measure $\mu_\La$ defined on these observables.  In the last section of
\cite{zoology} we illustrate how these mathematical objects are encountered
in Aerts' model system for a spin-${1\over 2}$ quantum entity.

\section{An axiomatics for context dependence.} 
 
In this section we translate the assumptions of the previous section in an axiomatic way. 

\subsection{Measurement systems (m.s.).}

We characterize the physical entities that we consider by the following objects:  
\par\noindent $\bullet$ a set of states $\Si$ and a set of
measurements ${\cal E}$. 
\par\noindent $\bullet$ $\forall e\in {\cal E}$, a set of outcomes $O_e$ represented as a measurable
subset of the real line.     
\par\noindent $\bullet$ $\forall p\in\Si,\forall e\in {\cal E}$: a probability measure
$P_{p,e} : {\cal B}_e\rightarrow[0,1]$, where ${\cal B}_e$ are the measurable subsets of
$O_e$. 
\par\noindent We call $\Sigma,\cal E$ a m.s$.$ and denote the collection 
of all m.s$.$ as ${\bf MS}$.  
Let $O_{\cal E}=\cup_{e\in{\cal E}}O_e$, ${\cal B}_{\cal E}=\{B\in {\cal B}_e|e\in {\cal
E}\}$ and ${\cal P}_{\cal E}=\{B\subseteq O_e|e\in{\cal E}\}$.
For a fixed set of outcomes $O$ and a fixed set of states $\Sigma$, the set of all
$\Si,{\cal E}\in{\bf MS}$ with $O_{\cal E}\subseteq O$ is denoted as ${\bf
MS}(\Sigma,O)$. If ${\cal E}$ contains only one measurement $e$ we call it a {\it one
measurement system} (abbreviated as 1m.s$.$), and we denote it as $\Sigma,e$. The
collection of all 1m.s$.$ is denoted as ${\bf MS}_0$.    
To summarize all probability measures that characterize a m.s$.$ within one
mathematical object we introduce a map  
$P_{\Si,{\cal E}}:
\Sigma\times {\cal E}\times {\cal B}_{\cal E}\rightarrow [0,1]$,
which is such that $\forall p\in\Sigma,\forall e\in{\cal E}$:
$P_{p,e}$ is the trace of $P_{\Si,{\cal E}}$ for 
a restricted domain $\{p\}\times \{e\}\times {\cal
B}_e$, and for all $B\in{\cal B}_{\cal E}$: 
\beq\label{eq:PpetovOE}
P_{p,e}(B)=P_{p,e}(B\cap O_e)
\eeq
For this collection of m.s$.$ we express in the following definition the relation '$\ldots$
is representable as $\ldots$' in a mathematical way. 
\bdf\label{df:mathequi}
Two m.s$.$ $\Sigma,\cal
E$ and $\Sigma ',\cal E'$ are called mathematically equivalent (denoted by $\Sigma,{\cal
E}\sim\Sigma ',\cal E'$) if there exist two maps $\zeta: \Sigma \rightarrow \Sigma '$ and 
$\eta: \cal E \rightarrow \cal E'$, both one to one and
onto, and if $\forall e\in {\cal E}$, there exists a
$\sigma$-isomorphism $\nu:{\cal B}_e\rightarrow {\cal B}_{\eta(e)}$
such that: 
\beq
\forall p \in\Sigma,\forall B \in {\cal B}_e: P_{p,e}(B)= P_{\zeta(p),\eta (e)}(\nu(B))
\eeq 
\edf
\noindent Clearly, theorems
on the existence of certain representations of a m.s$.$ can be expressed in
terms of mathematical equivalence. 
We end this section the notion of 'belonging up to  mathematical
equivalence'. 
Let $\Sigma,{\cal E}\in {\bf MS}$ and
${\bf N},{\bf N}'\subseteq {\bf MS}$. If there exists
$\Sigma ',\cal E'\in {\bf N}$ such that $\Sigma ',{\cal E}'\sim\Sigma,{\cal E}$ we
write:
\beq
\Sigma,{\cal E}\insim {\bf N}
\eeq
 
\subsection{Hidden measurement systems (h.m.s.).} 
 
In the following definition we introduce these m.s$.$ that are related to 
parameterized sets of 'compatible' strictly classical observables, i.e., strictly
classical observables with a common set of states and a common set of outcomes.
\bdf
$\Sigma,{\cal E}\in {\bf MS}$ is
called 'strictly classical' if $\forall e\in {\cal E}$, $e$ is a 'strictly
classical measurement', i.e., $\forall p \in \Sigma, \forall B\in {\cal B}_e:
P_{p,e}(B)\in
\{0,1\}$.
\edf 
\noindent If $\Si,{\cal E}$ is a strictly classical m.s$.$ then, $\forall e\in {\cal E}$ there
always exists a strictly classical observable 
$\varphi_e : \Sigma\rightarrow
O_e$
such that $\forall p \in \Sigma$ and $\forall B \in
{\cal B}_e$ we have $P_{p,e}(B)={\bf 1}_B[\varphi_e(p)]$ (${\bf 1}_B$ is the indicator\footnote{The
indicator ${\bf 1}_B:O_e\rightarrow\{0,1\}$ is such that $\forall o\in B:{\bf 1}_B(o)=1$
and 
$\forall o\in O_e\backslash B:{\bf 1}_B(o)=0$.} of $B$).
We use this property in the following definition, where we introduce a parameterization of
a set of strictly classical measurements with common sets of states and outcomes. In this
definition we denote
$P_{p,e_\la}$ as
$P_{p,\la}$ and the set of all subsets of the set $\La$ as ${\cal P}_\La$.    
\bdf\label{deflambda} 
Let ${\cal E}=\{ e_\lambda |
\lambda\in\Lambda\}$ and let $O_{\cal E}$ be the outcomes of $e_\la$ for all $\la\in\La$.
$\Sigma,{\cal E}\in {\bf MS}$ is called a '$\Lambda$-m.s.' if there exists a set
\beq
\Phi_\Lambda  =\{\varphi_\lambda: \Sigma\rightarrow O_{\cal E}|\lambda\in\Lambda\}
\eeq
which is such that $\forall p \in \Sigma,\forall\la\in\La,\forall B \in {\cal
B}_{\cal E}: P_{p,\la}(B)={\bf 1}_B[\varphi_\lambda(p)]$.
We introduce a map
$\Delta\Lambda : \Sigma\times {\cal P}_{\cal E}\rightarrow {\cal P}_\La$  
such that
$\forall p\in\Sigma,\forall o\in O_{\cal E},\forall B\in {\cal P}_{\cal E}: 
\Delta\Lambda_p^o =\{\lambda\in\Lambda |\varphi_\lambda
(p)=o\}$ and $\Delta\Lambda_p^B = \cup_{o\in B}\Delta\Lambda_p^o$
($\Delta\Lambda_p^o$
is the image of $(p,\{ o\})$
and $\Delta\Lambda_p^B$ the image of $(p,B)$). 
\edf
\noindent One easily verifies that we are able to restrict the
domain of $\Delta\La$ to $\Si\times {\cal B}_{\cal E}$. 
To avoid notational overkill, we apply the same notations for the map $\Delta\La$ when
defined on $\Si\times {\cal B}_{\cal E}$ as  when defined on $\Si\times {\cal P}_{\cal E}$
(which of the two domains we consider will follow from the context, or will be specified). For a
fixed state $p\in\Si$, we can consider 
$\Delta\La_p:{\cal B}_{\cal E}\rightarrow{\cal
P}_\La$, i.e$.$, $\Delta\La$ with the domain restricted to $\{p\}\times {\cal
B}_{\cal E}$. 
For every fixed state $p\in\Si$ we can introduce
$\varphi_p:\La\rightarrow O_{\cal E}$ which is such that  
$\forall \la\in\La: \varphi_p(\la)=\varphi_\la(p)$.  
Let $\DL (\Sigma\times {\cal B}_{\cal E}) = \{\DL_p^B|p \in \Sigma,\forall B \in
{\cal B}_{\cal E}\}$.    
\bpr\label{pr:DL=si-mor}
Let ${\cal B}_\La$ be a sub-$\si$-field of ${\cal P}_\La$ and let $\DL (\Sigma\times
{\cal B}_{\cal E})\subseteq {\cal B}_\La$.
For all $p\in\Si$, $\Delta\La:\Si\times {\cal B}_{\cal E}\rightarrow{\cal
B}_\La$ defines a $\si$-morphism, namely $\Delta\Lambda_p:{\cal B}_{\cal
E}\rightarrow{\cal B}_\La$, and $\forall p\in\Si$, $\varphi_p:\La\rightarrow O_{\cal
E}$ is a measurable function.  
\epr
\noindent The proof of this proposition is straightforward and therefore omitted.    

In the following definition we introduce a probability measure on a collection of strictly
classical observables in the following sense: we consider a new (in general non-classical)
measurement by supposing that one of the strictly classical measurements corresponding with
the strictly classical observables occurs with a given probability.  The idea of defining new
measurements by performing one measurement in a collection has been introduced by
Piron (see \cite{pir76} and \cite{pir90}).  
The idea of creating non-classical measurements by
considering classical measurements, equipped with a relative frequency of 
occurrence, has been
introduced by Aerts in his model system for a spin-${1\over2}$ quantum entity
(see \cite{aer86} and \cite{aer94}), which was based on the model in
\cite{gispir}.
\bdf\label{def:hmsyst}
A '$\Lambda$-hidden measurement model' $\Sigma, {\cal
E},\mu_\La$ consists in: 
\par\noindent i) a $\Lambda$-measurement system $\Sigma, {\cal E}$
\par\noindent ii) a probability measure $\mu_\La : {\cal B}_\La\rightarrow [0,1]$
that fulfills $\DL (\Sigma\times {\cal
B}_{\cal E})\subseteq {\cal B}_\La$
\par\noindent Define $e_\mu$ as the
measurement which is such that a strictly classical measurement
$e_\la\in{\cal E}$ occurs with the probability determined by $\mu$, i.e$.$, $\forall B\in{\cal
B}_\La$, the probability that $\la\in B$ is
$\mu_\La(B)$.  
The 1m.s$.$ $\Sigma,e_\mu$ related to $\Sigma, {\cal
E},\mu_\La$ is called a '$\mu_\La$-h.m.s$.$'. If $\mu_\La$ is not
specified, but $\La$ is, we call it a '$\La$-h.m.s$.$'. If $\mu_\La$ nor
$\La$ are specified, we call it a 'h.m.s$.$' 
\edf
\noindent Thus, every $\Lambda$-hidden measurement model defines a new one
measurement system if we suppose that $\mu_\La$ expresses a lack of knowledge concerning which
$e_\lambda\in {\cal E}$ actually takes place. Since in general, the measurements $e_\mu$ 
are not strictly classical, they are related to non-classical
observables. 
In this definition one easily sees that $\La$ can indeed be interpreted as the
set of states of the measurement context in the sense that for every given
$\la\in\La$, $e_\la$ determines an interaction between the entity under study and the
measurement context.
\bpr\label{pr:DLisO.K.}
Let $\Sigma, e_\mu$ be the 1m.s$.$ related to a $\Lambda$-hidden measurement model $\Sigma, {\cal
E},\mu_\La$ and let $P_{p,e_\mu}$ be the trace of $P_{\Si,e_\mu}$ for a restricted domain
$\{p\}\times \{e_\mu\}\times {\cal B}_{\cal E}$. $\forall p\in\Sigma, \forall B\in {\cal
B}_{\cal E}$:
\beq
P_{p,e_\mu}(B) =\mu_\La (\Delta\Lambda_p^B)
\eeq 
\epr 
\bpf
Since $\DL (\Sigma\times {\cal B}_{\cal E})\subseteq {\cal B}_\La$, $P_{\Si,e_\mu}$ is well
defined:
\[ \begin{array}{ccc}
 & _{P_{\Si,e_\mu}} & \\
 \Sigma\times {\cal B}_{\cal E}\ & \longrightarrow & \ [0,1] \\
 & & \\
\ \ \ _{\DL}\searrow & & \nearrow_{\mu_\La}\ \ \\
 & {\cal B}_\La & 
\end{array} \]
$\forall p\in\Sigma,\forall B
\in {\cal B}_{\cal E}$: 
$P_{p,e_\mu}(B)=\mu_\La (\{ \la | \varphi_\la (p)\in B\})
=\mu_\La (\DL_p^B)$. 
\epf
\noindent Define the set of all h.m.s$.$ in ${\bf MS}_0$ as ${\bf HMS}_0$, the set of all
$\La$-h.m.s$.$ in ${\bf MS}_0$ as ${\bf HMS}_0(\La)$, and the set of all
$\mu_\La$-h.m.s$.$ in ${\bf MS}_0$ as ${\bf HMS}_0(\mu_\La)$. In the following definition we
extend Definition \ref{def:hmsyst} to h.m.s$.$ with multiple
non-classical measurements, all of them defined in the same way as we defined $e_\mu$ in
Definition \ref{def:hmsyst}, i.e., we suppose that $\forall e\in {\cal E}$, there exists a
set of classical observables, paramertized by a set $\Lambda$ of states of the measurement
context.
\bdf\label{df:HMSgen}
Let $\Si,{\cal E}\in {\bf MS}$. If $\forall e\in {\cal E}:\Si,e\in {\bf
HMS}_0$ we call
$\Si,{\cal E}$ a h.m.s.\footnote{We remark that the symbol ${\cal E}$ 
which appears in Definition \ref{def:hmsyst} (i.e., a $\La$-set of strictly classical
measurements) is from a conceptual point of view completely different from the one which appears
in Definition \ref{df:HMSgen} (any set of measurements on an entity with $\Si$ as set of
states such that all $e\in{\cal
E}$ are defined in the same way as we defined $e_\mu$ in Definition \ref{def:hmsyst}), i.e.,
for every
$e\in {\cal E}$ of Definition
\ref{df:HMSgen} there exists a set of strictly classical measurements ${\cal E}_e$.}. If
$\forall e\in {\cal E}:\Si,e\in {\bf HMS}_0(\La)$ we call
$\Si,{\cal E}$ a $\La$-h.m.s. If $\forall e\in {\cal E}:\Si,e\in {\bf
HMS}_0(\mu_\La)$ we call
$\Si,{\cal E}$ a $\mu_\La$-h.m.s.
\edf  
\noindent The set of all h.m.s$.$ is denoted as ${\bf HMS}$.  For a fixed set $\Lambda$, we
denote the set of all 
$\La$-h.m.s$.$ as ${\bf HMS}(\La)$. For a fixed probability measure $\mu_\Lambda$, we denote the
set of all 
$\mu_\La$-h.m.s$.$ as ${\bf HMS}(\mu_\La)$ (when the
specification of $\La$ is not relevant, we will also use the simplified notation ${\bf
HMS}(\mu)$). Clearly we have ${\bf HMS}(\mu_\La)\subset {\bf HMS}(\La)\subset
{\bf HMS}\subset {\bf MS}$.
For a fixed set of states $\Sigma$ and a fixed set of outcomes $O$ we denote the set of all
h.m.s$.$ in ${\bf MS}(\Sigma,O)$ as
${\bf HMS}(\Sigma,O)$.  Again for fixed sets $\Sigma$ and $O$ we denote the set of all
$\La$-h.m.s$.$ in ${\bf MS}(\Sigma,O)$ as ${\bf HMS}(\Sigma,O,\Lambda)$ and
the set of  all $\mu_\La$-h.m.s$.$ in 
${\bf MS}(\Sigma,O)$ as ${\bf HMS}(\Sigma,O,\mu_\La)$. 
For every $\Si,{\cal
E}\in {\bf HMS}$ we can define a map $\DL:\Si\times{\cal
E}\times {\cal B}_{\cal E}\rightarrow {\cal P}_\La$, 
such that $\forall e\in {\cal E}$, the restriction of this new map to
$\Si\times\{e\}\times {\cal B}_e$ corresponds with the map introduced in Definition
\ref{deflambda} and, such that $\forall B\in
{\cal B}_{\cal E}:
\DL_{p,e}^B=\DL_{p,e}^{B\cap O_e}$ (we denote the restriction of this new
map to
$\{p\}\times\{e\}\times {\cal B}_{\cal E}$ as $\DL_{p,e}$). The
results of this section remain valid for this new map if $\forall e\in{\cal E}$, we replace
$\DL$ by the map $\DL_e:\Si\times {\cal B}_{\cal E}\rightarrow {\cal P}_\La$ (which is obtained by
restriction of the domain of $\DL:\Si\times{\cal E}\times{\cal B}_{\cal E}\rightarrow {\cal
P}_\La$),  if we replace $\DL_p$ by $\DL_{p,e}:{\cal B}_{\cal E}\rightarrow {\cal P}_\La$ and if
we replace $\varphi_p$ by $\varphi_{p,e}:\La\rightarrow O_{\cal E}$.
 
\subsection{Compatibility of the definition of a h.m.s$.$ with the geometric structure of quantum
mechanics.}\label{subsec:comp}

If there exists an additional structure on the
set of all possible outcomes of a measurement system\footnote{For example, a partial ordering of
the subsets of all outcomes and/or the implementation of spatial symmetries.}, one
could demand that this additional structure induces a structure on $\La$. 
In this section we show how the additional structure in the description of a physical entity can be
implemented in a straightforward way within this framework.  We consider the case of a 
quantum
entity submitted to measurements with a finite number of outcomes.  We will show that it suffices
to have a h.m.s.-representations for only one of the measurements to obtain a
representation for all measurements.
If ${\cal E}$ consists of all measurements with $n$ outcomes, we can represent such a measurement
by $n$ eigenvectors $p_{e,1},\ldots, p_{e,n}$ and $n$ corresponding  eigenvalues $o_{e,1},\ldots,
o_{e,n}$. 
Consider one given measurement $e_0$ (with $p_{0,1},\ldots, p_{0,n}$ as eigenvectors and
$o_{0,1},\ldots, o_{0,n}$ as respective eigenvalues) for which we have a h.m.s.-representation,
i.e., there exist:
\beq
\Phi_{\Lambda,0}=\{\varphi_{0,\lambda}:\Sigma\rightarrow\{p_{0,1},\ldots,
p_{0,n}\}|\lambda\in\Lambda\}
\eeq
and
\beq
\mu_{\Lambda,0}:{\cal B}_\Lambda\rightarrow [0,1] 
\eeq  
that characterize this h.m.s.-representation.  Then, we can define a representation for every
$e\in{\cal E}$ in the following way:
\beq\label{eqaerts1}
\Phi_{\Lambda,e}=\{\varphi_{e,\lambda}:\Sigma\rightarrow\{p_{e,1},\ldots, p_{e,n}\}:
p\mapsto U_e\circ\varphi_{0,\lambda}\circ U_e^{-1}(p)|\lambda\in\Lambda\}
\eeq
and
\beq\label{eqaerts2}
\mu_{\Lambda,e}=\mu_{\Lambda,0}    
\eeq  
where $U_e$ is the unitary transformation defined by $\forall i:p_{e,i}=U_e(p_{0,i})$.  In this way,
the h.m.s.-representation clearly 'respects' the structure that characterizes this quantum 
entity.  For an example of the application of eq.\ref{eqaerts1} and eq.\ref{eqaerts2} we refer
to Aerts' model system which can be found in \cite{aer86}, \cite{aer91}, 
\cite{aer95d}, \cite{coe95c} and \cite{thesisbob}, and which is also discussed
within the formalism of this approach in \cite{zoology}.

\section{On the existence of h.m.s.-representations.}

Before we proceed we need to introduce some measure theoretical notations and lemma's. 
Nonetheless, to avoid a notational overkill in the main section of this paper, we have
collected all lemma's and proofs in an appendix at the end of this
paper.

\subsection{Some mathematical preliminaries and notations.}\label{sec:math}
 
First we will introduce and study a
collection of mathematical objects that'll play a crucial role in the characterization of the
h.m.s$.$ in ${\bf HMS}$, and thus, also in the criterion for the existence of
h.m.s.-representations which will be presented at the end of this section.
\bdf\label{def:measure sp} 
Let ${\cal B}$ be a Borel algebra, and let $\mu:{\cal
B}\rightarrow [0,1]$ be a probability measure. Define ${\cal B}/\mu$ as the set of equivalence
classes for the relation $\sim$ on ${\cal B}$, which is defined by: $B\sim
B'\Leftrightarrow\mu(B\triangle B')=0$. 
We call $({\cal B},\mu)$
a measure space if ${\cal B}\cong{\cal B}/\mu$, i.e.:
\beq
\{ B | B\in {\cal B},\mu(B)=0\}=\{\emptyset\}
\eeq 
Two measure spaces $({\cal B},\mu)$ and $({\cal B'},\mu')$ are isomorphic (denoted as $({\cal
B},\mu)\cong ({\cal B'},\mu')$), if there exists a $\si$-isomorphism $H:{\cal B}\rightarrow {\cal
B'}$ which is such that $\forall B\in {\cal B}: \mu(B)=\mu'(H(B))$.
\edf
\noindent One can verify that ${\cal B}/\mu$ is again a Borel algebra, and that $\mu$
induces a probability measure on ${\cal B}/\mu$. For a proof we refer to
\cite{bir40}.
The Borel sets of $[0,1]$ will
be denoted by ${\cal B}_{[0,1]}$ and the Lebesgue
measure by $\mu_{[0,1]}$. 
The quotient
${\cal B}_{[0,1]}/\mu_{[0,1]}$ is denoted by ${\cal
B}_{\Bbb R}$ and the probability measure introduced on  
${\cal B}_{\Bbb R}$ by $\mu_{[0,1]}$ as 
$\mu_{\Bbb R}$. If we consider the measure space $({\cal B}_{\Bbb R},\mu_{\Bbb R})$, 
we omit the index ${\Bbb R}$ in $\mu_{\Bbb R}$ (in Lemma 1 we will
see that that this cannot lead to any confusion).     To characterize 'not to big' Borel
algebras we have the following definition:  
\bdf  
We call a Borel algebra ${\cal B}$ separable 
if there exists a countable dense subset, i.e$.$,
if there exists a set ${\cal D}=\{B_i | i\in {\Bbb N}\}$ which is such that the
smallest Borel subalgebra of ${\cal B}$ containing ${\cal D}$ is ${\cal B}$ itself.
We call a measure space $({\cal B},\mu)$ separable if ${\cal B}$ is separable.
\edf
\noindent Let ${\bf M}$ be the collection of all classes consisting of isomorphic separable
measure spaces, i.e., every ${\cal M}$ in ${\bf M}$ is a class of isomorphic separable
measure spaces. In the appendix at the end of this paper, we characterize ${\bf M}$ in an
explicit way. On
${\bf M}$ we introduce the following relation\footnote{In
\cite{zoology} we prove that ${\bf M},\leq$ is a poset, i.e., $\leq$ is a partial order
relation.}.
\bdf\label{def:lsim}
Define a binary relation $\leq$ on ${\bf M}$ by: ${\cal M}\leq {\cal M}'$
if $\forall ({\cal B},\mu)\in {\cal M}$ and $\forall ({\cal B}',\mu')\in
{\cal M}'$, there exists a $\si$-morphism $F: {\cal B}\rightarrow {\cal B}'$ such
that $\forall B\in {\cal B}:\mu'(F(B))=\mu (B)$.
\edf
\noindent Clearly, it suffices to have one $\si$-morphism $F$ such that
$\forall B\in{\cal B}:\mu'(F(B))=\mu (B)$. 
\bpr\label{onetoone}  
The $\si$-morphism $F$ in Definition \ref{def:lsim} is one to
one.
\epr 
\noindent The proof of this proposition is straightforward and omitted. 
Denote the set of all
integers, smaller or equal then a given
$n\in{\Bbb N}$ as ${\Bbb X}_n$.  Let ${\cal B}_n$ be the Borel algebra of
all subsets of ${\Bbb X}_n$ and let ${\cal B}_{{\Bbb N}}$ be the Borel algebra of
all subsets of ${\Bbb N}$. Denote the class of all sets isomorphic 
with ${\Bbb X}_n$ as ${\bf X}_n$, the class of all sets isomorphic
with ${\Bbb N}$ as ${\bf X}_{{\Bbb N}}$, 
and the class of all sets isomorphic
with ${\Bbb R}$ as ${\bf X}_{\Bbb R}$. 
Let ${\bf X}=\cup_{n\in{\Bbb N}}{\bf X}_n\cup{\bf X}_{{\Bbb N}}\cup{\bf
X}_{\Bbb R}$. 
For a given set $X\in{\bf X}$, denote the set of all subsets of $X$ as ${\cal
P}_X$. There exists a one-to-one map 
$h_X:X\rightarrow [0,1]$, and thus, we can consider 
${\cal B}_{X,{\Bbb R}}=\bigl\{\{x|h_X(x)\in B\}\bigm|B\in {\cal
B}_{[0,1]}\bigr\}\subseteq {\cal P}_X$. 
Clearly, $h_X$ is a measurable function, i.e., we can consider the
$\si$-morphism $H_X:{\cal B}_{[0,1]}\rightarrow {\cal B}_{X,{\Bbb R}}$ induced by
this measurable function.
Let ${\bf MX}$ be the
collection of all triples  $(X,{\cal B}_X,\mu_X)$, where $X\in {\bf X}$,
${\cal B}_X={\cal B}_{X,{\Bbb R}}$ and
$\mu_X:{\cal B}_X\rightarrow [0,1]$ is a probability measure. 
In the following proposition we prove a connection between the relation $\leq$ on
${\bf M}$ and the existence of measurable functions for objects in ${\bf MX}$.
\bpr\label{th:fieldchar2} 
Let $(X,{\cal B}_X,\mu_X)$ and $(Y,{\cal
B}_Y,\mu_Y)$ in ${\bf MX}$, and suppose that the measure space related to ${\cal B}_X$
and $\mu_X$ belongs to ${\cal M}_X$, and the one related to ${\cal
B}_Y$ and $\mu_Y$ belongs to ${\cal M}_Y$.
If ${\cal M}_X\leq {\cal M}_Y$,
there exists a measurable function $f:Y\rightarrow X$ such
that the related $\si$-morphism $F:{\cal B}_{X}\rightarrow {\cal B}_{Y}$ fulfills $\forall
B\in {\cal B}_X: \mu_X(B)=\mu_Y(F(B))$.
\epr
For the proof of this proposition we refer to the appendix at the end of this paper.

\subsection{A criterion on the existence of h.m.s.-representations.}

In this section we identify an explicit criterion that enables us to verify whether a given
h.m.s$.$ is a representation of a given m.s.  
This criterion will be the main key in the proof on
the existence for a h.m.s.-representation for every m.s. Moreover, as it has been shown in
\cite{thesisbob} and \cite{zoology}, this criterion also enables us to build a complete
classification of all possible h.m.s.-representations for a given quantum-like m.s. 
Nonetheless, in this paper we only want to show that our definition for context dependence
can be imposed on every m.s.

If no confusion is possible, we write $\mu\in
{\bf MX}$ (or
$\mu_\La\in {\bf MX}$) in stead of $(\La, {\cal B}_\mu,\mu_\La)\in {\bf MX}$. 
Consider $\Si,{\cal E}\in{\bf MS}$ with an
event probability $P_{\Si,{\cal E}}:\Si\times {\cal E}\times {\cal B}_{\cal E}\rightarrow [0,1]$.
$\forall p\in\Si,\forall e\in {\cal E}$ we denote ${\cal B}_e/P_{p,e}$ as ${\cal
B}_{p,e}$, and the induced probability measure on ${\cal B}_{p,e}$ as
$\mu_{p,e}$. $\forall \Si,{\cal E}\in{\bf MS}$, $({\cal
B}_{p,e},\mu_{p,e})$ is a separable measure space
for all $p\in\Si$ and for all $e\in {\cal E}$, and thus, $(O_e, {\cal B}_e, P_{p,e})\in{\bf
MX}$. 
\bit
\item Let ${\cal M}_{p,e}$ be the unique class in ${\bf M}$ such that $({\cal
B}_{p,e},\mu_{p,e})\in{\cal M}_{p,e}$. 
\item $\forall\Si,{\cal E}\in{\bf MS}$ we introduce: 
$\Delta{\bf M}(\Si,{\cal E})=\{{\cal M}_{p,e}|p\in\Si,e\in {\cal E}\}$
\eit
For every $\Si,e\in {\bf HMS}_0$ there exists $\mu_\La$ such that
$\Si,e\in {\bf HMS}_0(\mu_\La)$. Denote ${\cal B}_\La/\mu_\La$ as ${\cal
B}_\mu$, and the induced probability measure on ${\cal B}_\mu$ as $\mu$. 
Analogously, if $\Si,{\cal E}\in {\bf HMS}$, we can define ${\cal
B}_\mu,\mu$ for all $e\in {\cal E}$. For $\Si,{\cal E}\in {\bf HMS}(\mu_\La)$, there
exists one unique measure space $({\cal B}_\mu,\mu)$, which is called 'the measure space
related to the
$\mu_\La$-h.m.s$.$ $\Si,{\cal E}$'. For $\Si,{\cal E}\in{\bf HMS}(\La)$, we
have to consider a measure space 
$({\cal B}_\mu,\mu)$ for all $e\in {\cal E}$.
\bit
\item Let
${\cal M}_\mu$ be the unique class in ${\bf M}$ such that $({\cal B}_\mu,\mu)\in{\cal
M}_\mu$.   
\eit
For a h.m.s$.$ in
$\Si,{\cal E}\in{\bf HMS}(\La)$ we have to consider one measure space 
${\cal B}_\mu,\mu$ for all $e\in {\cal E}$. For every $\La\in {\bf X}$ we 
introduce the following
subset of ${\bf M}$: 
\bit  
\item ${\bf M}_\La=\{{\cal M}_\mu|\mu_\La\in{\bf MX}\}$ 
\eit 
We also introduce the following relation on subsets of ${\bf M}$.
\bdf
$\forall {\bf N},{\bf N}'\subseteq {\bf M}$: 
\beqa
{\bf N}\leq {\bf N}'\Longleftrightarrow\forall {\cal M}\in {\bf N},  \exists {\cal M}'\in {\bf
N}':{\cal M}\leq{\cal M}'
\eeqa 
\edf
\noindent We'll denote ${\bf
N}\leq \{{\cal M}\}$ as ${\bf N}\leq {\cal M}$ and $\{{\cal M}\}\leq {\bf N}$ as ${\cal
M}\leq {\bf N}$.
In the following definition we introduce a subcollection of ${\bf HMS}$ that contains these
h.m.s$.$ in which appear only separable measure spaces.
\bdf
Let ${\bf HMS}^S_0$ be the collection of all $\Si,e\in{\bf HMS}_0$ such that $({\cal B}_\mu,\mu)$ is a
separable measure space and let ${\bf HMS}^S$ be the collection of all $\Si,{\cal E}\in{\bf HMS}$
such that $\forall e\in{\cal E}$: $\Si,e\in{\bf HMS}_0^S$.
\edf
\noindent In the following section, we will prove that it suffices to consider measure spaces
contained in classes in ${\bf M}$, and this automatically allows us to
limit ourselves to h.m.s$.$ in ${\bf HMS}^S$. 

Now we identify  the necessary and sufficient condition for the existence of a
$\mu_\Lambda$-h.m.s$.$-representation in ${\bf HMS}(\Si,O_{\cal E},\mu_\La)$, for a given
m.s$.$ in ${\bf MS}$.
\bth\label{th:rep1} 
Let $\Si, {\cal E}\in {\bf MS}$ and $\mu_\La\in {\bf MX}$:
\beq
\Si,{\cal E}\insim {\bf HMS}(\Si,O_{\cal E},\mu_\La)\Leftrightarrow\Delta{\bf M}(\Si,{\cal E})\leq
{\cal M}_\mu
\eeq 
\eth
\bpf
$\Longrightarrow$ Let $e\in{\cal E}$. According to Definition \ref{df:HMSgen}, there exists
$\Si,{\cal E}',\mu_\La$ such that
$\Si,e\sim\Si,e_\mu$. Thus, there exists a
$\si$-morphism  $\nu:{\cal B}_e\rightarrow{\cal B}_{{\cal E}'}$ which is such that  $\forall B
\in {\cal B}_e: P_{p,e_\mu}(\nu(B))=P_{p,e}(B)$ ($\zeta :\Si\rightarrow\Si$ is the
identity, $\eta:\{e\}\rightarrow\{e_\mu\}$ is trivial).  Moreover, there exists
$\Delta\La_{p,e}:{\cal B}_{\cal E}\rightarrow{\cal B}_\La$ (see Proposition
\ref{pr:DL=si-mor}) which is such that  $\forall B \in {\cal B}_{{\cal E}}:
\mu_\La(\DL_{p,e}(B))=P_{p,e_\mu}(B\cap O_e)$ (see Proposition \ref{pr:DLisO.K.}). Since
${\cal B}_{\cal E'}\subseteq{\cal B}_{\cal E}$, we can consider
the map 
$[\DL_{p,e}\circ\nu]: {\cal B}_e\rightarrow{\cal B}_\La$. Clearly, $[\DL_{p,e}\circ\nu]$
is also a $\si$-morphism and fulfills  $\forall B \in {\cal B}_e:
\mu_\La([\DL_{p,e}\circ\nu](B))=P_{p,e}(B)$. Define $F_{p}:{\cal B}_e\rightarrow {\cal
B}_{p,e}$ and $F_\mu:{\cal B}_\La\rightarrow {\cal B}_\mu$ by the following scheme: 
\par\vspace{-2pt}\par\noindent  
\[  
\begin{array}{ccccccc}
& & _{_\nu} &  & _{_{\DL_{p,e}}} & &\\
& {\cal B}_e & \longrightarrow & {\cal B}_{\cal E'} &
\longrightarrow & {\cal B}_\La &\\ & \hspace{-2mm}^{_{F_{p}}}\downarrow\ 
& _{_{P_{p,e}}}\hspace{-3.5mm}\searrow\ \ & ^{_{P_{p,e_\mu}}}\hspace{-2mm}\downarrow
\ \ \ \ & \ \ \swarrow\hspace{-2mm}_{_{\mu_\La}} & \ \ \downarrow ^{_{F_\mu}} &\\  
& {\cal B}_{p,e} & \longrightarrow & [0,1] & \longleftarrow & {\cal B}_\mu &\\ 
& & ^{_{\mu_{p,e}}} &  & ^{_{\mu}} & &\\
\end{array} 
\] 
\par\vspace{-0pt}\par\noindent
Thus, $\forall B \in {\cal B}_e: 
\mu([F_\mu\circ\DL_{p,e}\circ\nu](B))=P_{p,e}(B)$. 
For all $B\in{\cal B}_{p,e}$, there exists at least
one $B_1\in{\cal B}_e$ such that $F_p(B_1)=B$.   
Let $B_1'=[F_\mu\circ\DL_{p,e}\circ\nu](B_1)\in{\cal B}_\mu$. 
If $B_2\not=B_1$ and $F_p(B_2)=B$, then $P_{p,e}(B_1\triangle B_2)=0$, and thus 
\beqa
\mu([F_\mu\circ\DL_{p,e}\circ\nu](B_1)\triangle
[F_\mu\circ\DL_{p,e}\circ\nu](B_2))&=&\\
\mu([F_\mu\circ\DL_{p,e}\circ\nu](B_1\triangle B_2))&=&\\
P_{p,e}(B_1\triangle B_2)&=&0
\eeqa 
By definition of $F_\mu$ there exists only one
$B_1'=[F_\mu\circ\DL_{p,e}\circ\nu](B_2)=[F_\mu\circ\DL_{p,e}\circ\nu](B_1)$.
Thus,  we can define $F_\nu:{\cal B}_{p,e}\rightarrow {\cal B}_\mu$ such that 
$\forall B\in{\cal B}_{p,e}:
F_\nu(B)=[F_\mu\circ\DL_{p,e}\circ\nu](B')\Leftrightarrow B=F_{p}(B')$.
\[ 
\begin{array}{ccccccc}
& & _\nu &  & _{\DL_{p,e}}& &\\
& {\cal B}_e & \longrightarrow & {\cal B}_{\cal E'} & \longrightarrow & {\cal B}_\La &\\
& & & & & &\\
& \ \ _{F_{p}}\searrow & & _{F_\nu} & & \swarrow_{F_\mu}\ \ & \\ 
& &{\cal B}_{p,e} & \longrightarrow & {\cal B}_\mu & &\\ 
\end{array} 
\]
Let $B'\in {\cal B}_e$ be
such that 
$F_\nu(B)=[F_\nu\circ F_{p}](B')$. We have,
$\mu(F_\nu(B))=\mu([F_\nu\circ F_{p}](B'))=\mu([F_\mu\circ\DL_{p,e}\circ\nu](B'))=
\mu_\La([\DL_{p,e}\circ\nu](B'))=P_{p,e_\mu}(\nu(B'))=P_{p,e}(B')=\mu_{p,e}(B)$,
and thus, Definition \ref{def:lsim} is fulfilled.
As a consequence,
${\cal M}_{p,e}\leq {\cal M}_{\mu}$, and thus, $\Delta{\bf
M}(\Si,{\cal E})\leq {\cal M}_\mu$. 
 
\noindent $\Longleftarrow$ Let
$p\in\Si$ and $e\in {\cal E}$. Since ${\cal M}_{p,e}\leq {\cal M}_\mu$, and, since
both $(O_e, {\cal B}_e, P_{p,e})$ and $(\La,{\cal B}_\La, \mu_\La)$ are in ${\bf MX}$, we can
apply Proposition \ref{th:fieldchar2}. Thus, there exists a
measurable function $f_p:\La\rightarrow O_e$ such that the related $\si$-morphism
$F_p:{\cal B}_{e}\rightarrow {\cal B}_{\La}$ fulfills $\forall B\in {\cal B}_e:
P_{p,e}(B)=\mu_\La(F_p(B))$. Define $\DL_e:\Si\times{\cal
B}_{\cal E}\rightarrow {\cal B}_\La$ such that $\forall B\in {\cal B}_{\cal
E}:\DL_{p,e}^B=F_p(O_e\cap B)$. Define $\varphi_\la:\Si\rightarrow X$ such that
$\forall p\in\Si$: $\vlp=f_p(\la)$. We have $\forall p\in\Si$:
$\DL_p^B=\{\la|\la\in\La, f_p(\la)\}=\{\la|\la\in\La, \vlp\}$. Thus, there exists a
set of strictly classical observables ${\cal E}_e$. Thus, $\DL_e$ defines
a $\La$-m.s. Still following Proposition \ref{th:fieldchar2}, $\forall
B\in{\cal B}_e:P_{p,e}(B)=\mu_\La(F(B))$, and thus, $\forall B\in{\cal B}_{\cal
E}:P_{p,e}(B)=\mu_\La(F(O_e\cap B))=\mu_\La(\DL_{p,e}^B)$ (see eq.\ref{eq:PpetovOE}). If we
identify $e$ with $e_\mu$, the measurement related to $\Si,{\cal E}_e,\mu_\La$, we
obtain $\Si,e\insim {\bf HMS}_0(\Si,O_{\cal E},\mu_\La)$, and thus, $\Si,{\cal E}\insim {\bf
HMS}(\Si,O_{\cal E},\mu_\La)$. 
\epf 
An alternative version of this theorem expresses the
sufficient and necessary condition for the existence of at least one 
representation in ${\bf HMS}^S(\Si,O_{\cal E},\La)$:   
\bth\label{th:rep2}  
Let $\Si, {\cal E}\in {\bf MS}$ and $\La\in {\bf X}$: 
\beq
\Si,{\cal E}\insim {\bf HMS}^S(\Si,O_{\cal E},\La)\Leftrightarrow\Delta{\bf M}(\Si,{\cal E})\leq {\bf
M}_\La
\eeq 
\eth 
\bpf
We have $\Si,{\cal E}\insim{\bf HMS}^S(\Si,O_{\cal E},\La)\Leftrightarrow$
$\forall e\in{\cal E}:\Si,e\insim{\bf HMS}^S_0(\Si,O_{\cal E},\La)\Leftrightarrow$ 
$\forall e\in{\cal E},\exists \mu_\La:
\Si,e\insim{\bf HMS}_0(\Si,O_{\cal
E},\mu_\La)\Leftrightarrow$ $\forall e\in{\cal E},\exists
\mu_\La:\Delta{\bf M}(\Si,e)\leq {\cal M}_\mu\Leftrightarrow
\forall e\in{\cal E},\exists{\cal M}_\mu\in{\bf M}_\La:\Delta{\bf M}(\Si,e)\leq {\cal
M}_\mu\Leftrightarrow$ 
$\forall e\in{\cal E}:\Delta{\bf M}(\Si,e)\leq {\bf
M}_\La\Leftrightarrow$
$\Delta{\bf M}(\Si,{\cal E})\leq {\bf M}_\La$.
\epf 

\subsection{A proof for the existence of h.m.s.-representations for all m.s.}
 
In the following theorem we prove that the axiomatics for the dependence on
the measurement context imposed by the definition of a h.m.s$.$ implies no restriction
for a general m.s$.$, i.e., every m.s$.$ can be represented as a h.m.s., with $[0,1]$ as set
of states of the measurement context.
\bth\label{th:ext.theo.} 
$\forall\Si,{\cal E}\in{\bf MS}$: 
$\Si,{\cal E}\insim {\bf HMS}^S(\Si,O_{\cal E},[0,1])$.
\eth
\bpf 
According to Lemma \ref{lm:MleqMR} we know that ${\bf M}\leq {\cal M}_{\Bbb R}$.
For all $\Si,{\cal E}\in{\bf MS}$ we have $\Delta{\bf M}(\Si,{\cal E})\leq {\bf
M}$.  Thus, $\Delta{\bf
M}(\Si,{\cal E})\leq {\cal M}_{\Bbb R}$, and thus, ${\Si,{\cal E}}\insim {\bf
HMS}(\Si,O_{\cal E},\mu_{[0,1]})\subseteq{\bf HMS}^S(\Si,O_{\cal E},[0,1])$.
\epf 

\section{Conclusion.} 
   
Every m.s$.$ in ${\bf MS}$ has a representation as a h.m.s$.$ in
${\bf HMS}$, and thus, also quantum mechanics can be
represented in this way. As a
consequence,  the h.m.s.-formalism that is presented in this paper can be seen as {\it
an axiomatics for general physical entities for context dependence that leads to a
classical representation of non-classical systems.} We also identified the general condition for 
the existence of a h.m.s$.$-representation
with $\Lambda$ as set of 'states of the measurement context', or with $\mu_\Lambda$ as
relative frequency of occurence of these states of the measurement context.  If no further
restrictions or assumptions are made on
$\La$, we only obtain restrictions on the ordinality of
$\Lambda$, and on
the specific probability measure $\mu_\La$ that we consider. 
A lot of problems are still to be
solved, for example, how precisely should this h.m.s.-formalism be fitted in the more general
operational formalisms for quantum mechanics like Piron's approach (see \cite{pir76} and
\cite{pir90}) or the Foulis-Randall approach (see \cite{fou83} and
\cite{fou78}).  Still, we think that the approach
presented in this paper certainly leads to a successful extension of the
contemporary quantum framework as well from a philosophical as from a
mathematical point of view.

\section{Appendix: some measure theoretical lemma's.} 

\noindent Let ${\cal B}$ and ${\cal B}'$ be two Borel Algebras. Denote their direct
union\footnote{A more general construction, and also more details, can be found in
\cite{sik69}.} by ${\cal B}\duni
{\cal B}'$, i.e., ${\cal B}\duni
{\cal B}'=\{(B,B')|B\in
{\cal B}, B'\in {\cal B}'\}$ equipped with three relations:
\beqa
(B_1,B'_1)\cup (B_2,B'_2)&=&(B_1\cup B_2, B'_1\cup B'_2)\\
(B_1,B'_1)\cap (B_2,B'_2)&=&(B_1\cap B_2, B'_1\cap B'_2)\\
^c(B_1,B'_1)&=&(^cB_1,^cB'_1)
\eeqa
In the following definition we introduce an extension
of this notion of direct union of Borel algebras to the collection of measure
spaces, i.e., we introduce a way to 'compose' measure spaces.
\bdf  
Let $({\cal B},\mu)$ and $({\cal B'},\mu')$ be measure spaces, $a\in ]0,1[$ and $\mu\dsuma\mu' :
{\cal B}\duni{\cal B}'\rightarrow [0,1]$
such that $\forall (B,B')\in {\cal B}\duni
{\cal B}':\mu\dsuma\mu'(B,B')=(1-a)\mu(B)+a\mu'(B')$. Define the
weighted direct union $({\cal B},\mu)\dunia ({\cal B}',\mu')$ of $({\cal B},\mu)$ and
$({\cal B'},\mu')$ as the measure space\footnote{One easily verifies that this weighted
direct union is indeed a measure space.} $({\cal B}\duni {\cal B}',\mu\dsuma\mu')$.
\edf
\noindent As in section \ref{sec:math}, we denote the set of all
integers, smaller or equal then a given
$n\in{\Bbb N}$ as ${\Bbb X}_n$.  Let ${\cal B}_n$ be the Borel algebra of
all subsets of ${\Bbb X}_n$ and let ${\cal B}_{{\Bbb N}}$ be the Borel algebra of
all subsets of ${\Bbb N}$. We introduce the following sets of monotonous
decreasing strictly positive functions:
\beqa 
M_n&=&\{m: {\Bbb X}_{n}\rightarrow [0,1]|\sum_{i=1}^{i=n}m(i)=1, i\leq j\Rightarrow
m(j)\leq m(i)\}\\ M_{{\Bbb N}}&=&\{m: {\Bbb N}\rightarrow [0,1]|\sum_{i\in
{{\Bbb N}}}m(i)=1, i\leq j \Rightarrow m(j)\leq m(i)\}
\eeqa
\par\noindent For all $m\in M_n\cup M_{{\Bbb N}}$
we define a probability measure 
$\mu_m: {\cal B}_N\rightarrow [0,1]$ by
$\forall i:\mu_m(\{i\})=m(i)$. We also introduce the following notations for some classes of measure
spaces: 
\beqa
{\cal M}_{\Bbb R}=\{({\cal B},\mu)|({\cal B},\mu)\cong ({\cal B}_{\Bbb R},\mu)\}
\eeqa
\par\noindent $\forall N\in{\Bbb N}\cup\{{{\Bbb N}}\},\forall m\in M_N:$
\beqa
{\cal M}_N^m=\{({\cal B},\mu)|({\cal B},\mu)\cong ({\cal B}_N,\mu_m)\}
\eeqa
\par\noindent $\forall N\in{\Bbb N}\cup\{{{\Bbb N}}\},\forall m\in M_N, \forall
a\in]0,1[:$ 
\beqa
{\cal M}_{N,a}^m=\{({\cal B},\mu)|({\cal B},\mu)\cong ({\cal B}_{\Bbb R},\mu)\dunia
({\cal B}_N,\mu_m)\}
\eeqa 
\par\noindent and also the following notations for sets of such classes:  
\beqa 
&&{\bf M}_N=\bigl\{{\cal M}_N^m \bigm|m\in M_N\bigr\}\\
&&{\bf M}_{{\Bbb R},a}=\bigl\{{\cal M}_{N,a}^m \bigm|  N\in{\Bbb N}\cup\{{{\Bbb N}}\}, m\in M_N\bigr\}\\
&&{\bf M}=\cup_{N\in{\Bbb N}\cup\{{{\Bbb N}}\}}{\bf
M}_N\cup_{a\in]0,1[}{\bf M}_{{\Bbb R},a}\cup\{{\cal M}_{\Bbb R}\}
\eeqa
The use of this symbol ${\bf M}$ (which we used in section \ref{sec:math} as a notation for
the collection of all classes consisting of isomorphic separable measure spaces) is justified
by the following lemma.
\blm\label{th:Bunique} 
The collection of all separable measure spaces is:
\beq
{\cal M}_{\Bbb R}\cup_{N\in{\Bbb N}\cup\{{\Bbb N}\}}\cup_{m\in M_N}{\cal
M}_N^m\cup_{a\in]0,1[}{\cal M}_{N,a}^m
\eeq
Moreover, for every separable measure space $({\cal B},\mu)$, $\exists! {\cal
M}\in {\bf M}$ such that $({\cal B},\mu)\in {\cal M}$. 
\elm
\noindent The proof is a rather
long construction that uses Lemma \ref{lm:birk}, Lemma
\ref{pr:decompdunia}, Lemma \ref{lm:at most} (see further) and the Loomis-Sikorki theorem (see
\cite{loo47} and
\cite{sik48}). Since the content if the theorem agrees with our intuition,
and the proof of it doesn't contribute in an essential way to the understanding of the subject
of this paper, this proof is omitted. An explicit proof with the notations of this paper can
be found in
\cite{thesisbob}.
\blm\label{lm:birk} 
If ${\cal B}$ is a separable Borel algebra with 
$\{B\in {\cal B}| B'\subset B\Rightarrow B'=\emptyset\}=\{\emptyset\}$,
then ${\cal B}\cong {\cal B}_{\Bbb R}$.
Moreover, for every probability measure $\mu: {\cal B}\rightarrow [0,1]$, there exists a
$\si$-isomorphism $F_\mu: {\cal B}\rightarrow {\cal B}_{\Bbb R}$ such that $\forall B\in
{\cal B}: \mu (B)=\mu_{\Bbb R}\bigl(F_\mu(B)\bigr)$. 
\elm
\bpf 
This lemma is proved by Marczewski. For an outline of it we refer to \cite{bir40} or
\cite{hor48}. 
\epf
\blm\label{pr:decompdunia}
Let $({\cal B},\mu)$ be a measure space, $B_0\in {\cal B}$, $a=\mu(B_0)$, ${\cal
B}_l=\{B\in {\cal B}| B\cap B_0=\emptyset\}$ and ${\cal B}_r=\{B\in {\cal B}|
B\cap B_0=B\}$. Define two maps, $\mu_l: {\cal B}_l\rightarrow [0,1]$ and $\mu_r:
{\cal B}_r\rightarrow [0,1]$ such that $\forall B\in {\cal B}_l:\mu_l(B)={\mu(B)\over {1-a}}$
and $\forall B\in {\cal B}_r:\mu_r(B)={\mu(B)\over a}$.
Then, both $({\cal B}_l,\mu_l)$ and $({\cal
B}_r,\mu_r)$ are measure spaces. Moreover we have $({\cal B},\mu)\cong ({\cal
B}_l,\mu_l)\dunia ({\cal B}_r,\mu_r)$.
\elm
\bpf
One easily sees that ${\cal B}_r$ (resp. ${\cal B}_l$) are Borel algebras,
with $B_0$ (resp. $B_0^c$) as greatest element. By definition, $\mu_l$ and $\mu_r$ are
$\si$-additive. Since $\mu(B_0)=a$ and $\mu(B_0^c)=1-a$, both $\mu_l$ and $\mu_r$
are normalized. Thus, $\mu_l$ and $\mu_r$ are probability measures, and thus, $({\cal
B}_l,\mu_l)$ and $({\cal B}_r,\mu_r)$ are measure spaces. 
We have to show that there exists a $\si$-isomorphism $H: {\cal B}\rightarrow {\cal
B}_l\duni {\cal B}_r$ such that $\forall B\in {\cal B}, \forall (B_l,B_r)\in {\cal
B}_l\duni {\cal B}_r\, :\, (B_l,B_r)=H(B)\Rightarrow \mu(B)=\mu_l\dsuma\mu_r(B_l,B_r)$.
Since $\forall B\in {\cal B}$ we have:
$\mu_l\dsuma\mu_r(B\cap B_0^c,B\cap
B_0) =(1-a)\mu_l(B\cap B_0^c)+a\mu_r(B\cap B_0)
=\mu\bigl((B\cap B_0^c)\cup(B\cap B_0)\bigr)
=\mu(B)$, we can define $H$ by $\forall B\in {\cal B}:
H(B)=(B\cap B_0^c,B\cap B_0)$.
\epf
\blm\label{lm:at most}
A measure space cannot have an uncountable subset of disjoint elements with a
nonzero probability. 
\elm
\bpf
Suppose that there exists such a set ${\cal D}$. Let ${\cal D}_i=\{B|B\in {\cal D}, \mu(B)>{1\over
i}\}$. 
Clearly, ${\cal
D}=\cup_{i\in {{\Bbb N}}}{\cal D}_i$. Since ${\cal D}$ is uncountable, there exists
$n\in{\Bbb N}$ such that
${\cal D}_n$ contains an infinite set of elements. Let ${\cal
D}'_n=\{B_i|i\in {{\Bbb N}}\}$ be a countable subset of ${\cal D}_n$. We have
$\mu(\cup_{B\in {\cal D}_n})\geq\mu(\cup_{B\in {\cal D'}_n}) =\sum_{i\in {{\Bbb N}}}\mu(B_i) \geq\sum_{i\in {{\Bbb N}}}{1\over n}=\infty$.
\epf 
\blm\label{lm:fieldchar} 
Let $\mu_1:{\cal B}_{[0,1]}\rightarrow [0,1]$ and 
$\mu_2:{\cal B}_{[0,1]}\rightarrow [0,1]$ be two probability measures such that ${\cal
B}_{[0,1]}/\mu_1\cong{\cal B}_{[0,1]}/ \mu_2\cong{\cal B}_{\Bbb R}$.
There exists a measurable function $f:[0,1]\rightarrow [0,1]$, which is such that the related
$\si$-morphism $F:{\cal B}_{[0,1]}\rightarrow {\cal B}_{[0,1]}$ fulfills $\forall B\in {\cal
B}_{[0,1]}: \mu_1(B)=\mu_2(F(B))$.
\elm 
\bpf
Let $b\in [0,1]$. We prove that there exists $x\in [0,1]$ such that
$\mu_1([0,x])=b$. Suppose that $x$ doesn't exist.
Let $b_-$ be the supremum of all $b'\in [0,b[$ such
that there exists $x'\in [0,1]$ fulfilling $\mu_1([0,x'])=b'$. Then, there exists
an increasing sequence $(b_i)_i$ with for all $i\in{\Bbb N}$: $b_i\in
[b_--1/i,b_-]$  and $\exists x_i\in [0,1]$ such that $\mu_1([0,x_i])=b_i$.
Clearly, $b_-$ is the supremum of $\{b_i|i\in{\Bbb N}\}$ and $(x_i)_i$ is
also an increasing sequence. Denote the supremum of
$\{x_i|i\in{\Bbb N}\}$ as $x_-$. 
There are two possibilities $x_-\in\{x_i|i\in{\Bbb N}\}$ and
$x_-\not\in\{x_i|i\in{\Bbb N}\}$. 
If $x_-\in\{x_i|i\in{\Bbb N}\}$ then $\cup_{i\in{\Bbb N}}[0,x_i]=[0,x_-]$, and thus $\mu_1(\cup_{i\in{\Bbb N}}[0,x_i])=\mu_1([0,x_-])$. 
If $x_-\not\in\{x_i|i\in{\Bbb N}\}$ then $\cup_{i\in{\Bbb N}}[0,x_i]=[0,x_-[$, and
again we find
$\mu_1([0,x_-])=\mu_1([0,x_-[)=\mu_1(\cup_{i\in{\Bbb N}}[0,x_i])$, since 
$\mu_1(\{x_-\})=0$.  We also have
for all $i\in{\Bbb N}$: $\mu_1(]x_i,x_{i+1}])=\mu_1([0,x_{i+1}])-\mu_1([0,x_i])$. Thus: 
\beqa
\mu_1([0,x_-])&=&\mu_1(\cup_{i\in{\Bbb N}}[0,x_i])=
\mu_1([0,x_1]\cup(\cup_{i\in{\Bbb N}}]x_i,x_{i+1}]))\\
&=&\mu_1([0,x_1])+\sum_{i\in{\Bbb N}}\mu_1(]x_i,x_{i+1}])\\ 
&=&\mu_1([0,x_1])+\sum_{i\in{\Bbb N}}\mu_1([0,x_{i+1}])-
\sum_{i\in{\Bbb N}}\mu_1([0,x_i])\\
&=&b_1+\sum_{i\in{\Bbb N}}(b_{i+1}-
b_i)=b_-
\eeqa 
Define $b_+$ as the infimum of all $b'\in ]b,1]$ such
that $\exists x'\in [0,1]:\mu_1([0,x'])=b'$ (there exists at least one
such
$b'$ since $\mu_1([0,1])=1$).
Then, there exists an decreasing sequence $(b_i)_i$
with for all $i\in{\Bbb N}$: $b_i\in [b_+,b_++1/i]$ 
and $\exists x_i\in [0,1]$ such that $\mu_1([0,x_i])=b_i$. Denote the infimum of
$\{x_i|i\in{\Bbb N}\}$ as $x_+$. Clearly, 
$\cap_{i\in{\Bbb N}}[0,x_i]=[0,x_+]$ and $(x_i)_i$ is also an decreasing sequence. Thus:  
\beqa
\mu_1([0,x_+])&=&\mu_1(\cap_{i\in{\Bbb N}}[0,x_i])=\mu_1((\cup_{i\in{\Bbb N}}[0,x_i]^c)^c)\\  
&=&1-\mu_1(\cup_{i\in{\Bbb N}}]x_i,1])=
1-\mu_1(]x_1,1]\cup(\cup_{i\in{\Bbb N}}]x_{i+1},x_i]))\\ 
&=&1-(\mu_1(]x_1,1])+\sum_{i\in{\Bbb N}}\mu_1(]x_{i+1},x_i]))\\ 
&=&1-(1-\mu_1([0,x_1])+\sum_{i\in{\Bbb N}}(\mu_1([0,x_i])-\mu_1([0,x_{i+1}])))\\
&=&1-(1-b_1+\sum_{i\in{\Bbb N}}(b_i-b_{i+1}))=b_+
\eeqa  
For all $x'\in ]x_-,x_+[$ we have
$\mu_1([0,x'])\geq\mu_1([0,x_-])=b_-$, $\mu_1([0,x'])\leq\mu_1([0,x_+])=b_+$, but, 
as a consequence of the definition of $b_-$ and
$b_+$, there exist no $x'\in ]x_-,x_+[$ such that $\mu_1([0,x'])\in [b_-,b_+]$. Thus
we obtain a contradiction. As a consequence, $x$ exists.
For all $x\in [0,1]$, define $f$ such that $\mu_1([0,f(x)])=\mu_2([0,x])$. We can
define a $\si$-morphism  
$F:{\cal B}_{[0,1]}\rightarrow {\cal B}_{[0,1]}$ related to this measurable function.
Thus, $F([0,x])=\{y|f(y)\in [0,x]\}=\{y|f(y)\leq x\}=
\{y|\mu_2([0,y])\leq \mu_1([0,x])\}$ for all $x\in [0,1]$. For all $x_1,x_2\in [0,1]$ such that 
$x_1<x_2$: 
\beqa 
F(]x_1,x_2])&=&F([0,x_2]\setminus [0,x_1])=F([0,x_2])\setminus F([0,x_1])\\
&=&\{y|\mu_2([0,y])\leq \mu_1([0,x_2])\}\setminus\{y|\mu_2([0,y])\leq \mu_1([0,x_1])\}\\
&=&]y(x_1),y(x_2)]
\eeqa
where $y(x_1)$ is the smallest real in $[0,1]$ such that $\mu_2([0,y(x_1)])=\mu_1([0,x])$
and $y(x_2)$ is the largest real in $[0,1]$ such that $\mu_2([0,y(x_2)])=\mu_1([0,x])$.
All this leads us to
$\mu_2(F(]x_1,x_2]))=\mu_2(]y(x_1),y(x_2)])=\mu_2([0,y(x_2)])-\mu_2([0,y(x_1)])=
\mu_1([0,x_2])-\mu_1([0,x_1])=\mu_1(]x_1,x_2])$. By definition, ${\cal B}_{[0,1]}$
is the smallest Borel subalgebra of ${\cal P}_{[0,1]}$ containing
$\{]a,b]|0\leq a<b\leq 1; a,b\in[0,1]\}$. This completes the proof as a
consequence of the $\si$-additivity of $\mu_1$ and $\mu_2$.   
\epf 
\blm\label{lm:MleqMR}
${\bf M},\leq$ has a greatest ellement, namely ${\cal M}_{\Bbb R}$, i.e., ${\bf
M}\leq {\cal M}_{\Bbb R}$
\elm
\bpf
First we prove that $\forall {\cal M}_{{\Bbb N},a}^m\in {\bf
M}_{{\Bbb R},a}:{\cal M}_{{\Bbb N},a}^m\leq 
{\cal M}_{\Bbb R}$. Consider the
Borel algebra\footnote{One can easily prove that it poses no problem to extend the notion of
direct union to countable  sets of Borel algebras. For more details we refer to
\cite{sik69}.} $\duni_{i\in{\Bbb N}}{\cal B}_{\Bbb R}$, and a
probability measure $\mu': \duni_{i\in{\Bbb N}}{\cal B}_{\Bbb R}\rightarrow
[0,1]$ which is defined by the  relations
$\forall B\in {\cal B}_{\Bbb R}$ ($\mu$ is defined as in $({\cal B}_{\Bbb R},\mu)$):
$\mu'(B,\emptyset,\ldots)=(1-a).\mu(B);
\mu'(\emptyset,B,\emptyset,\ldots)=a.m(1).\mu(B);
\mu'(\emptyset,\emptyset,B,\emptyset,\ldots)=a.m(2).\mu(B);
\ldots$.  
One
verifies that $\{B\in\duni_{i\in{\Bbb N}}{\cal B}_{\Bbb R}|\mu(B)=0\}=\{\emptyset\}$ and that $\duni_{i\in{\Bbb N}}{\cal B}_{\Bbb R}$ is
separable, i.e., 
$\duni_{i\in{\Bbb N}}{\cal B}_{\Bbb R},\mu'$ is a separable measure space. 
Clearly,
there exists no $B\in\duni_{i\in{\Bbb N}}{\cal B}_{\Bbb R}$ with $\mu'(B)\not=0$,
and such that
$B'\in\duni_{i\in{\Bbb N}}{\cal B}_{\Bbb R}$ and $B'\subset B$ implies
$B'=\emptyset$, and thus, $(\duni_{i\in{\Bbb N}}{\cal B}_{\Bbb R},\mu')\cong ({\cal B}_{\Bbb R},\mu)$ (see Lemma \ref{lm:birk}), i.e., there exists a $\si$-isomorphism
$H:\duni_{i\in{\Bbb N}}{\cal B}_{\Bbb R}\rightarrow {\cal B}_{\Bbb R}$ such that $\forall B\in\duni_{i\in{\Bbb N}}{\cal B}_{\Bbb R}: \mu'(B)=\mu(H(B))$.  For all $B\in {\cal B}_{{\Bbb N}}$, define a map
$X_B:{\Bbb N}\rightarrow\{\emptyset, I\}$ which is such that $\forall i\in B:X_B(i)=I$ and $\forall i\not\in
B:X_B(i)=\emptyset$. We define a map  $F:{\cal B}_{\Bbb R}\duni {\cal
B}_{{\Bbb N}}\rightarrow\duni_{i\in{\Bbb N}}{\cal B}_{\Bbb R}$ by the
relations $\forall B\in {\cal B}_{\Bbb R}:F(B,\emptyset)=(B,\emptyset,\emptyset,\ldots)$ and  $\forall B\in {\cal B}_{{\Bbb N}}:F(\emptyset, B)=(\emptyset,
X_B(1), X_B(2), X_B(3),\ldots)$. One verifies that the $\si$-morphism $H\circ F:{\cal
B}_{\Bbb R}\duni {\cal B}_{{\Bbb N}}\rightarrow {\cal B}_{\Bbb R}$ fulfills the
requirements of Definition \ref{def:lsim} and thus we have  ${\cal M}_{{\Bbb N},a}^m\leq 
{\cal M}_{\Bbb R}$. Along the same lines one proves that $\forall {\cal
M}_{n,a}^m\in {\bf M}_{{\Bbb R},a}:
{\cal M}_{n,a}^m\leq  {\cal M}_{\Bbb R}$
and that ${\bf M}_{{\Bbb N}}\cup_{n\in{\Bbb N}} {\bf M}_{n}\leq {\cal
M}_{\Bbb R}$. As a consequence ${\bf M}\leq {\cal M}_{\Bbb R}$. 
\epf
\noindent We end this appendix with the proof of proposition \ref{th:fieldchar2}.
\par\medskip\noindent\bpf
Consider two $\si$-epimorphisms  
$F_X: {\cal B}_{X}\rightarrow{\cal B}_{X}/\mu_X$ and 
$F_Y: {\cal B}_{Y}\rightarrow{\cal B}_{Y}/\mu_Y$, which induce a
probability measure $\mu:{\cal B}_{X}/\mu_X\rightarrow [0,1]$, respectively 
$\mu':{\cal B}_{Y}/\mu_Y\rightarrow [0,1]$. Clearly, $({\cal B}_{X}/\mu_X,\mu)$ and 
$({\cal B}_{Y}/\mu_Y,\mu')$ are measure spaces.
There also exists
$F':{\cal B}_{X}/\mu_X\rightarrow {\cal B}_{Y}/\mu_Y$ which fulfills Definition
\ref{def:lsim}. Let ${\cal D}_X=\{B\in {\cal B}_X|\mu_X(B)\not=0, B\supset B'\in
{\cal B}_X\Rightarrow B'=\emptyset\}$. Since ${\cal D}_X$ is at most countable (see
Lemma
\ref{lm:at most}), there exists a smallest set
${\Bbb X}\in\cup\{{\Bbb X}_i|i\in{\Bbb N}\}$ of indices such that
${\cal D}_X=\{B_i|i\in{\Bbb X}\}$. $\forall i\in{\Bbb N}$: let $B'_i\in{\cal B}_Y$
be such that $F_Y(B'_i)=[F'\circ F_X](B_i)$, and $B''_i=B'_i\backslash(\cup_{j=1}^{j=i-1}B'_j)$.
Clearly, $\cup_{i\in{\Bbb X}_N}B''_i=
\cup_{i\in{\Bbb X}_N}B'_i$ and $\forall i,j\in{\Bbb X}:i\not=j\Rightarrow B''_i\cap B''_j=\emptyset$. 
Since $\forall i,j\in{\Bbb X}:i\not=j\Rightarrow B_i\cap B_j=\emptyset$, we have 
$\forall i\in{\Bbb X}:B_i\cap(\cup_{j=1}^{j=i-1}B_j)=\emptyset$, and thus,
$F_Y(B'_i)\cap(\cup_{j=1}^{j=i-1}F_Y(B'_j))=\emptyset$. As a
consequence, $\forall i\in{\Bbb X}:\mu_Y(B'_i\cap(\cup_{j=1}^{j=i-1}B'_j))=
\mu'(F_Y(B'_i)\cap(\cup_{j=1}^{j=i-1}F_Y(B'_j)))=0$, and thus, 
$\forall i\in{\Bbb X}:F_Y(B''_i)=F_Y(B'_i\backslash(\cup_{j=1}^{j=i-1}B'_j))=F_Y(B'_i)=[F'\circ
F_X](B_i)$, what leads to $\mu_Y(B''_i)=\mu'(F_Y(B''_i))=\mu'[F'\circ
F_X](B''_i))= \mu_X(B_i)$.  
Define 
$X_1=\cup_{i\in{\Bbb X}}B_i$, $X_2=X\backslash X_1$,
$Y_1=\cup_{i\in{\Bbb X}}B''_i$ and $Y_2=Y\backslash Y_1$.
Suppose
that $\mu_X(X_2)=\mu_Y(Y_2)\not=0$. Consider ${\cal B}'_X=\{X_2\cap B|B\in
{\cal B}_X\}$ and  ${\cal B}'_Y=\{Y_2\cap B|B\in {\cal B}_Y\}$. 
Following Lemma \ref{lm:birk} and Lemma
\ref{pr:decompdunia}, we know that ${\cal B}'_X/\mu'_X\cong{\cal B}'_Y/\mu'_Y\cong{\cal
B}_{\Bbb R}$ ($\mu'_X$ and $\mu'_Y$ are the restrictions of $\mu_X$
to ${\cal B}'_X$, respectively $\mu_Y$ to ${\cal B}'_Y$, multiplied by $1/\mu_Y(Y_2)$,
and thus, they correspond with $\mu_r$ in Lemma \ref{pr:decompdunia}). This
observation, together with the definition of ${\bf MX}$, leads to ${\cal B}'_X\cong{\cal
B}'_Y\cong{\cal B}_{[0,1]}$.  
Let $f:Y\rightarrow X$ be such that $\forall i\in{\Bbb X},\forall y\in B''_i: f(y)\in B_i$. 
There are two possibilities: $\mu_Y(Y_2)=0$ or $\mu_Y(Y_2)\not=0$.
If $\mu_Y(Y_2)=0$, $\forall y\in Y_2$: we can choose $f(y)$ in $X_2$.  
If $\mu_Y(Y_2)\not=0$, we define $f(y)$ for all $y\in Y_2$ by
applying Lemma \ref{lm:fieldchar} (i.e$.$, we identify ${\cal B}'_X$ and ${\cal B}'_Y$
with ${\cal B}_{[0,1]}$, $\mu'_X$ with $\mu_1$ and $\mu'_Y$ with
$\mu_2$).  We can define
the related $\si$-morphism $F:{\cal B}_{X}\rightarrow {\cal B}_{Y}$. We find that 
$\forall i\in{\Bbb X}:F(B_i)=B''_i$, what leads to $\mu_Y(F(B_i))=\mu_Y(B''_i)=\mu_X(B_i)$.
$\forall B\in{\cal B}'_Y: \mu_Y(F(B))=\mu_X(B)$, as a consequence of 
Lemma \ref{lm:fieldchar}.
\epf
    
\section{Acknowledgments.}     
 
We thank Prof. D. Aerts for the discussions on the idea of hidden
measurements and we thank Dr. A. Wilce, Prof. K. E. Hellwig and the referee for their usefull
suggestions.  This work is supported by the IUAP-III n$^{\rm o}$9. The author is Research
Assistant of the National Fund for Scientific Research.

\end{document}